\documentclass[aps,prl,twocolumn,superscriptaddress,amsmath,amssymb,floatfix]{revtex4-2}
\usepackage[T1]{fontenc}
\usepackage[utf8]{inputenc}
\usepackage{graphicx,bm}
\usepackage[dvipsnames]{xcolor}
\DeclareRobustCommand{\changed}[1]{{\color{blue}#1}}
\usepackage[colorlinks=true,linkcolor=blue!60!black,citecolor=blue!60!black,urlcolor=blue!60!black]{hyperref}

\newcommand{\dd}{\mathrm{d}}
\newcommand{\T}{\mathsf{T}}
\newcommand{\G}{\mathsf{G}}
\newcommand{\Sv}{\mathbf{S}}
\newcommand{\Id}{\mathbf{1}}
\newcommand{\mH}{\mathcal{H}}
\DeclareMathOperator{\Rea}{Re}
\newcommand{\FP}{\mathrm{FP}}
\newcommand{\arxiv}[1]{\href{https://arxiv.org/abs/#1}{arXiv:#1}}

\begin{document}

\title{Modular Hamiltonian and modular flow of a massive scalar in an interval}

\author{Ra\'ul E. Arias}
\email{rarias@fisica.unlp.edu.ar}
\affiliation{Instituto de F\'isica La Plata -- CONICET and Departamento de F\'isica,\\
Universidad Nacional de La Plata, C.C. 67, 1900 La Plata, Argentina}
\author{Daniel Fondevila}
\affiliation{Instituto de F\'isica La Plata -- CONICET and Departamento de F\'isica,\\
Universidad Nacional de La Plata, C.C. 67, 1900 La Plata, Argentina}

\begin{abstract}
We obtain the vacuum modular Hamiltonian of a free massive scalar in an interval, for any $mL$. Its eigenvectors are superpositions of massive plane waves emitted from the two end points, with amplitudes given by a linear integral equation whose Neumann series is an expansion in reflections, and a normalization fixed by a Painlev\'e V transcendent. We follow the modular flow of initially localized wave packets inside the causal diamond. The flow is not geometric: the commutator profiles spread and can develop peaks ahead of the conformal trajectory.
\end{abstract}

\maketitle

\textit{Introduction.}---The vacuum state reduced to a region $V$ can be written as $\rho_V\propto e^{-\mH}$, where $\mH$ is the modular Hamiltonian. It generates the modular flow $\rho_V^{-i\tau}O\rho_V^{i\tau}$ of the operators localized in $V$ \cite{Haag}, and in recent years it has become a central object in the study of entanglement in quantum field theory (QFT). It gives the relative entropy between states \cite{Casini:2008cr,Blanco:2013joa}, it enters the proofs of energy conditions \cite{Balakrishnan:2017bjg,Ceyhan:2018zfg}, and in holography it is the key to bulk reconstruction \cite{Jafferis:2015del}. However, explicit modular Hamiltonians are scarce. When the region has enough symmetry the modular flow is geometric and $\mH$ is local: this is the case of the Rindler wedge in any QFT \cite{Bisognano:1975ih,Bisognano:1976za} and of spheres in conformal field theories \cite{Hislop:1981uh,Casini:2011kv}. More generally, the leading short-distance contribution to $\mH$ is local, with an inverse temperature fixed by the geometry \cite{Arias:2016nip,Arias:2017dda,Agon:2023tdb}. The nonlocal terms are known only in a few conformal examples \cite{Casini:2009vk,Longo:2009mn,Wong:2018svs,Arias:2018tmw,Fries:2019torus,Blanco:2019torus,Blanco:2019jhep}.

A mass breaks conformal symmetry. Small-mass expansions of
the modular Hamiltonian are known for free fermions
\cite{Cadamuro:2023xxs}. For the scalar field on an interval,
Ref.~\cite{Baranov:2024rdm} gives an exact reduced density
matrix and obtains the modular Hamiltonian in the small-
and large-mass limits. Its explicit spectral construction
at arbitrary $mL$ remained open. Numerical approximations of the modular operator and of the modular Hamiltonian of the double cone \cite{Bostelmann:2022yvh,Cadamuro:2023dc}, numerical studies of the massive scalar on the half line \cite{Minz:2025hl}, and lattice entanglement Hamiltonians \cite{Eisler:2020lyn} provide complementary information. The entanglement entropies of massive free fields in an interval have been known for twenty years \cite{Casini:2005rm,Casini:2005zv,Casini:2009sr}. They only need the spectrum of the correlator kernels, which is controlled by a Painlev\'e V transcendent but the modular Hamiltonian also needs the eigenvectors. In \cite{Arias:2017dda} it was shown that they satisfy linear differential equations whose coefficients involve the same transcendent, but the corresponding connection problem was left open.

\begin{figure*}[t]
\centering\includegraphics[width=\textwidth]{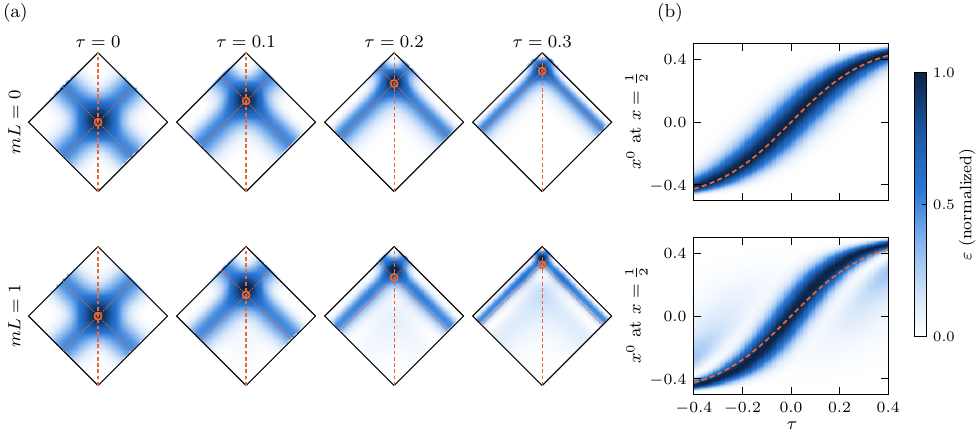}
\caption{Modular flow of $O=\int_A\gamma_1\phi$, centered at $x=\frac12$, with $x(1-x)\gamma_1=e^{-\omega^2/2\sigma^2}$, $\sigma=0.7$, for $mL=0$ (top) and $mL=1$ (bottom). (a) The energy density $\varepsilon$, eq.~\eqref{eq:eps}, of $\Phi_\tau(y)=i[\phi(y),O(\tau)]$ in the causal diamond $|x^0|<\min(x,1-x)$, normalized at $\tau=0$; $\Phi_\tau$ vanishes where $\phi(y)$ commutes with $O(\tau)$, and a point-like operator gives two null beams crossing at its position. Dashed: conformal orbit of the center; circle: conformal position at that $\tau$; dotted: null lines through it. (b) $\varepsilon$ on the axis $x=\frac12$ against $\tau$ and $x^0$, normalized at each $\tau$: the world line of the operator. Dashed: the conformal world line $x^0=\frac12\tanh(\pi\tau)$.}
\label{fig:flow}
\end{figure*}

In this Letter we solve this problem for the free massive scalar in an interval. We write the eigenvectors as superpositions of massive plane waves emitted from the two end points, with amplitudes given by a linear integral equation. This gives the modular Hamiltonian for any value of $mL$, and allows us to describe its nonlocal structure. With it we compute the modular flow of a massive field inside the causal diamond of the interval. The flow is not geometric: localized operators move faster than in the conformal case and they spread (Fig.~\ref{fig:flow}).

\textit{The problem.}---We take the interval $A=(0,1)$, measuring lengths in units of $L$, and we call $t=mL$. With $\phi(x)$ and $\Pi(x)$ the field and its conjugate momentum, the vacuum correlators $X(x,y)=\langle\phi(x)\phi(y)\rangle$ and $P(x,y)=\langle\Pi(x)\Pi(y)\rangle$ on $A$ are
\begin{equation*}
 X=\frac1{2\pi}K_0(t|x-y|),\qquad P=-\frac t{2\pi}\,\frac{K_1(t|x-y|)}{|x-y|}.
\end{equation*}
Since the state is Gaussian \cite{Peschel:2003rdm,Casini:2009sr},
\begin{equation}
 \mH=\tfrac12\!\int_A\!\!\dd x\!\int_A\!\!\dd y\big[\Pi(x)N_t(x,y)\Pi(y)+\phi(x)M_t(x,y)\phi(y)\big].
 \label{eq:H}
\end{equation}
As in \cite{Arias:2018tmw}, everything reduces to solving
\begin{equation}
 X\psi_s=\nu(s)\,\chi_s,\qquad P\chi_s=\nu(s)\,\psi_s,
 \label{eq:pair}
\end{equation}
with $\nu(s)=\frac12\coth(\pi s)$, where $s>0$ and $2\pi s$ is the modular energy. There are two solutions for each $s$ ($k=1,2$). Normalized as $\int_A\dd x\,\chi_s^k(x)^*\psi_{s'}^{k'}(x)=\delta_{kk'}\delta(s-s')$, they give
\begin{equation}
 \begin{aligned}
 N_t(x,y)&=\sum_k\int_0^\infty\!\!\dd s\,2\pi s\,\chi^k_s(x)\chi^k_s(y)^*,\\
 M_t(x,y)&=\sum_k\int_0^\infty\!\!\dd s\,2\pi s\,\psi^k_s(x)\psi^k_s(y)^*.
 \end{aligned}
 \label{eq:NM}
\end{equation}
For the chiral currents of the massless theory the eigenvectors are plane waves $e^{\pm is\omega(x)}$ in the variable $\omega(x)=\log\frac x{1-x}$, the two chiralities decouple, and one recovers the local conformal kernels 
\begin{equation}
N_0=\beta(x)\delta(x-y),\qquad M_0=-\partial_x[\beta(x)\partial_x\delta(x-y)],
\end{equation} 
with $\beta(x)=2\pi x(1-x)$ \cite{Hislop:1981uh,Casini:2011kv}.

Eqs.~\eqref{eq:pair} are equivalent to a problem in the Euclidean plane, parametrized by $z=x_1+ix_2$ and cut along $A$, \cite{Arias:2017dda,Arias:2018tmw}. We look for $S(z)$ with $(-\Delta+t^2)S=0$ outside $A$, $S\to0$ at infinity, bounded at the end points, and
\begin{equation}
 S^+=\lambda S^-,\qquad (\partial_2S)^+=\lambda(\partial_2S)^-\quad\text{on }A,
 \label{eq:jump}
\end{equation}
with $\lambda=e^{-2\pi s}$, where $\pm$ are the boundary values from above and below. Since $X$ and $P$ are the traces on the real line of the Green function $\frac1{2\pi}K_0(t|z|)$ and of its second derivative in $x_2$, Green's identity shows that $\chi=S^+$ and $\psi=(\partial_2S)^+$ solve \eqref{eq:pair}. Near each end point\changed{,} $S$ is a superposition of two waves $x^{\pm is}=e^{\pm is\log x}$, which give the continuous spectrum and the normalization.

\textit{The eigenvectors.}---Without mass\changed{,} $S$ is holomorphic and \eqref{eq:jump} is solved with elementary functions \cite{Arias:2018tmw}. The mass destroys holomorphy. Its role is taken by the massive plane waves 
\begin{equation}
e_\kappa(z)=\exp[-\frac t2(\kappa z+\bar z/\kappa)],
\end{equation}
which solve the Euclidean Klein--Gordon equation for every $\kappa\neq0$. For $\kappa=e^y>0$, $e_\kappa(x)=e^{-tx\cosh y}$ on the real line: it is a wave of rapidity $y$. We find that the eigenvectors are
\begin{equation}
 \begin{aligned}
 \Sv_s^+(x)&=e^{-\pi s}\big(I_x[\kappa^{is}p_s]-I_{1-x}[\kappa^{-is}h_s]\big),\\
 (\partial_2\Sv_s)^+(x)&=-\tfrac{it}2e^{-\pi s}\big(I_x[\kappa_-\kappa^{is}p_s]+I_{1-x}[\kappa_-\kappa^{-is}h_s]\big),
 \end{aligned}
 \label{eq:modes}
\end{equation}
where 
\begin{equation}
\kappa_-=\kappa-\kappa^{-1},\qquad
I_x[f]=\int_0^\infty\frac{\dd\kappa}\kappa\,e^{-\frac{tx}2(\kappa+\kappa^{-1})}f(\kappa), 
\end{equation}
and the amplitudes $p_s(\kappa),h_s(\kappa)\in\mathbb C^2$ are the solution of
\begin{equation}
 \begin{aligned}
 p_s(\kappa)&=e_1+i\,\frac{\sinh\pi s}\pi\int_0^\infty\!\dd\kappa'\,\frac{\kappa'^{-2is}\,d_t(\kappa')\,h_s(\kappa')}{\kappa+\kappa'},\\
 h_s(\kappa)&=e_2-i\,\frac{\sinh\pi s}\pi\int_0^\infty\!\dd\kappa'\,\frac{\kappa'^{2is}\,d_t(\kappa')\,p_s(\kappa')}{\kappa+\kappa'},
 \end{aligned}
 \label{eq:integral}
\end{equation}
with $d_t(\kappa)=e^{-\frac t2(\kappa+\kappa^{-1})}$, $e_1=(1,0)^T$ and $e_2=(0,1)^T$. The two components of $\Sv_s=(S^{(1)}_s,S^{(2)}_s)$ are two independent solutions. The term $I_x$ is a superposition of waves centered at the left end point, and $I_{1-x}$ of waves centered at the right one.

\begin{figure}[t]
\centering\includegraphics[width=\linewidth]{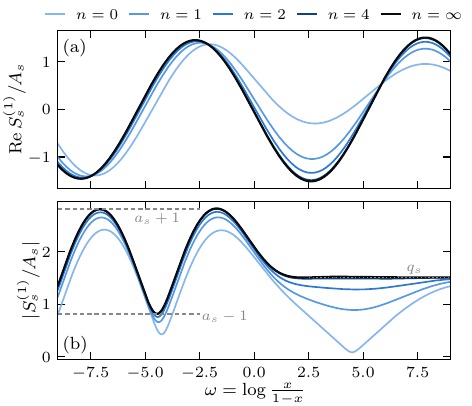}
\caption{The eigenvector built reflection by reflection: $S^{(1)}_s(x)/A_s$ for $s=0.6$ and $mL=0.3$, with the Neumann series of \eqref{eq:integral} started at $p_s=e_1$, $h_s=e_2$ (the Rindler functions of the two end points) and truncated after $2n+2$ iterations, i.e.\ after $n+1$ round trips between the end points, and the exact solution ($n=\infty$). The logarithmic coordinate $\omega(x)$ resolves the oscillatory behavior near both end points. (a) Real part. (b) Modulus. The dashed lines are the amplitudes $a_s\pm1$ and $q_s$ predicted by \eqref{eq:tips}.}
\label{fig:modes}
\end{figure}

Eq.~\eqref{eq:integral} has a simple physical meaning. At zeroth order, $p_s=e_1$ and $h_s=e_2$, and since $I_x[\kappa^{is}]=2K_{is}(tx)$ we get
\begin{equation*}
 S^{(1)}_s\simeq2e^{-\pi s}K_{is}(tx),\qquad S^{(2)}_s\simeq-2e^{-\pi s}K_{is}(t(1-x)).
\end{equation*}
These are the eigenfunctions of the Rindler (Bisognano--Wichmann) modular Hamiltonian of each end point. Each iteration brings a factor $d_t(\kappa')=e^{-t\cosh y'}$, which is the propagation of a wave of rapidity $y'$ across the whole interval. The Neumann series of \eqref{eq:integral} is then an expansion in reflections between the end points, see Fig.~\ref{fig:modes}. For every $s\neq0$ and $t>0$, the series converges geometrically, with rate bounded by $\tanh^2(\pi s)$ per round trip. Eq.~\eqref{eq:integral} is the Cauchy representation of the solution of a Riemann--Hilbert problem in the rapidity plane, whose jumps are chosen such that \eqref{eq:modes} has exactly the cut \eqref{eq:jump} on $A$ and no cut outside (End Matter).

\textit{End points and Painlev\'e V.}---The normalization of the eigenvectors, and with it $\mH$, is fixed at the end points. All the information about them is contained in
\begin{equation}
 \T_s=\big(p_s(0),h_s(0)\big)=\begin{pmatrix}a_se^{i\delta_s}&-iq_s\\ iq_s&a_se^{-i\delta_s}\end{pmatrix},
\end{equation}
with $q_s(t)$ and $\delta_s(t)$ real and $a_s=\sqrt{1+q_s^2}$. Near the end points, for $x\to0^+$ and $\varrho=1-x\to0^+$,
\begin{equation}
 \begin{aligned}
 \Sv_s^+(x)&\simeq A_s\,e_1\,x^{-is}+B_s\,(\T_s)_1\,x^{is},\\
 \Sv_s^+(1-\varrho)&\simeq-B_s\,e_2\,\varrho^{is}-A_s\,(\T_s)_2\,\varrho^{-is},
 \end{aligned}
 \label{eq:tips}
\end{equation}
where $A_s=e^{-\pi s}\Gamma(is)(t/2)^{-is}$, $B_s=A_s^*$, and $(\T_s)_k$ is the $k$-th column of $\T_s$. For $S_s^{(1)}(z)$, a wave incident from the left is reflected back to the left and transmitted to the right with amplitude magnitudes $1/a_s$ and $|q_s|/a_s$, respectively. The matrix $\T_s\in SU(1,1)$ transfers the wave coefficients between the two end points. For $t\to\infty$, $q_s\to0$ and the end points decouple; for $t\to0$, $q_s\to\infty$ and the waves are transmitted, as for the chiral modes. The dependence on $t$ is isomonodromic \cite{Jimbo:1981zz}, and gives ($'=\partial_t$)
\begin{equation}
 \begin{aligned}
 q_s''+\frac{q_s'}t&=\frac{q_s(q_s')^2}{1+q_s^2}+q_s(1+q_s^2)-\frac{4s^2q_s}{t^2(1+q_s^2)},\\
 t\,\delta_s'&=-\frac{2s\,q_s^2}{1+q_s^2},
 \end{aligned}
 \label{eq:PV}
\end{equation}
with the limits
\begin{equation*}
 \begin{aligned}
 q_s&\simeq2s\big[\log\tfrac2t-2\gamma_E-\Rea\psi(is)\big], &&t\to0,\\
 q_s&\simeq\tfrac{2\sinh\pi s}\pi K_{2is}(t), &&t\to\infty.
 \end{aligned}
\end{equation*}
This is the Painlev\'e V equation of the entropies \cite{Casini:2005zv,Casini:2009sr}, and the function $d(t)$ introduced in \cite{Arias:2017dda} is exactly $1/q_s$. The phase $\delta_s$ alone gives all the entanglement and R\'enyi entropies, which reproduce the known results \cite{Casini:2005zv}, and the capacity of entanglement studied in \cite{Arias:2023cap} (see End Matter).

\textit{The modular Hamiltonian.}---Since $X$ and $P$ are symmetric, eigenvectors with different $s$ are orthogonal, and their normalization follows from \eqref{eq:tips} (see End Matter). Then \eqref{eq:NM} becomes
\begin{equation}
 \begin{aligned}
 N_t(x,y)&=\int_0^\infty\!\!\dd s\,n_s(x,y),\\
 n_s(x,y)&=\frac{s\sinh\pi s}{\pi a_s}\,\hat\Sv_s(x)^T\begin{pmatrix}a_s&-iq_se^{-i\delta_s}\\ iq_se^{i\delta_s}&a_s\end{pmatrix}\hat\Sv_s(y)^*,
 \end{aligned}
 \label{eq:kernels}
\end{equation}
where $\hat\Sv_s=e^{\pi s}\Sv^+_s$. $M_t$ is given by the same formula with $e^{\pi s}(\partial_2\Sv_s)^+$ in place of $\hat\Sv_s$. Eqs.~\eqref{eq:modes}, \eqref{eq:integral} and \eqref{eq:kernels} give the modular Hamiltonian \eqref{eq:H} for any $t>0$. The kernels are distributions, and their invariant content is in the quadratic forms 
\begin{equation}
N_t[\gamma,\gamma]=\sum_{k}\int_0^\infty\!\dd s\,2\pi s\,\left|\int_A\!\dd x\,\chi_s^k(x)^*\gamma(x)\right|^2,
\end{equation}
which give the relative entropy of coherent states \cite{Casini:2019qst}. Completeness follows from the resolvent and is confirmed numerically (see End Matter).

The structure of $\mH$ follows from the behavior of the spectral densities at large $s$, where the eigenvectors approach the conformal ones,
\begin{equation}
 \begin{aligned}
 n_s(x,y)&=2\cos\vartheta_-+\sqrt{\tfrac{2t}s}\,(x+y-1)\sin\vartheta_++O(s^{-1}),\\
 \vartheta_-&=s[\omega(x)-\omega(y)]-t(x-y),\\
 \vartheta_+&=s[\omega(x)+\omega(y)]-t(x+y-1).
 \end{aligned}
 \label{eq:nsUV}
\end{equation}
The first term gives the universal local contribution,
\begin{equation*}
 \mH_{\rm loc}
 =\frac12\int_A\dd x\,\beta(x)
 \big[\Pi^2+(\partial_x\phi)^2\big].
\end{equation*}
Its coefficient is independent of the mass, in agreement with the universality of local temperatures \cite{Arias:2016nip}. The remaining terms give the nonlocal couplings, and also mass-dependent local terms, as in the Bisognano--Wichmann limit below. Near the diagonal the leading nonlocal terms are of order $t$:
\begin{equation}
 \begin{aligned}
 N_t&\simeq\beta(x)\delta(x-y)+2t\beta(x)X_0(x-y),\\
 M_t&\supset2t\beta(x)P_0(x-y),
 \end{aligned}
\end{equation}
where $X_0$ and $P_0$ are the massless correlators,
\begin{equation*}
 X_0=-\frac1{2\pi}\log|x-y|,\qquad P_0=-\frac1{2\pi(x-y)^2}.
\end{equation*} The second term of \eqref{eq:nsUV} mixes the two chiralities
and singles out the mirror line $y=1-x$. Since its prefactor vanishes there, the next order is needed. For fixed $x\neq\frac12$ and $y=1-x+\varsigma$, $\varsigma\to0$, the terms in $\vartheta_+$ are
\begin{equation}
n_s\supset\sqrt{\frac{2t}s}\Big[\varsigma\sin\vartheta_++\frac{3x(1-x)}{2s}\cos\vartheta_+\Big],\;\;\vartheta_+\simeq\frac{s\varsigma}{x(1-x)}.
 \label{eq:mirrorterm}
\end{equation}
Integrated over $s$, both give a cusp, the second three times larger and of opposite sign:
\begin{equation}
 N_t(x,1-x+\varsigma)-N_t(x,1-x)\simeq-2\sqrt{\pi t\,x(1-x)}\,|\varsigma|^{1/2},
 \label{eq:mirror}
\end{equation}
while 
\begin{equation}
M_t\simeq-\frac12\sqrt{\pi tx(1-x)}\,\FP|\varsigma|^{-3/2},
\end{equation}
where $\FP$ is the Hadamard finite part, 
\begin{equation}
\langle\FP|\varsigma|^{-3/2},f\rangle=\lim_{\epsilon\to0}\big[\int_{|\varsigma|>\epsilon}|\varsigma|^{-3/2}f\,\dd\varsigma-4\epsilon^{-1/2}f(0)\big].\nonumber 
\end{equation}
$N_t$ has a cusp and $M_t$ a homogeneous singularity of degree $-\frac32$. This coupling does not exist without mass, where the chiralities decouple; Fig.~\ref{fig:kernel} shows it on the lattice.

Two limits complete the picture. For $t\to\infty$, $\T_s\to\Id$, and the Kontorovich--Lebedev inversion formula gives, near each end point, the Bisognano--Wichmann kernels of the half line,
\begin{equation*}
 N_t\to2\pi x\,\delta(x-y),\qquad M_t\to2\pi\big[-\partial_x(x\partial_x\delta)+t^2x\,\delta\big],
\end{equation*}
where $\delta\equiv\delta(x-y)$. For $t\to0$ the mass enters the correlators only through a constant,
\begin{equation*}
 X=X_0+\tfrac1{2\pi}\big(\log\tfrac2t-\gamma_E\big)+O(t^2\log t),
\end{equation*}
while $P=P_0+O(t^2\log t)$. The constant contribution to $X$ is a rank one operator,
proportional to the projector on constant functions. We use this structure to obtain the leading correction to the modular Hamiltonian. To first order in $1/\Lambda$, with
$\Lambda=2(\log\frac8t-\gamma_E)$,
\begin{equation}
 \begin{aligned}
 N_t&=\beta\,\delta(x-y)+\tfrac\pi\Lambda\big[1-\max(|2x-1|,|2y-1|)\big],\\
 M_t&=-\partial_x\big[\beta\,\partial_x\delta(x-y)\big]+\tfrac{2\pi}\Lambda\big[\delta(x-y)+\delta(x+y-1)\big].
 \end{aligned}
 \label{eq:small}
\end{equation}
The momentum acquires a pyramid-shaped kernel, and the field a local term and a coupling between reflected points, as in \cite{Baranov:2024rdm}, with different coefficients (see End Matter below). At finite mass both deltas are smoothed at distances $\lesssim t\,x(1-x)$, the second into \eqref{eq:mirror}.

\textit{Modular flow.}---For a linear operator $O=\int_A(\gamma_1\phi+\gamma_2\Pi)$, the flow $O(\tau)=e^{i\tau\mH}Oe^{-i\tau\mH}$ keeps this form, with
\begin{equation*}
 \partial_\tau\gamma_1=-M_t\gamma_2,\qquad \partial_\tau\gamma_2=N_t\gamma_1.
\end{equation*}
The eigenvectors diagonalize it: each mode only acquires the phase $e^{2\pi is\tau}$ (End Matter). To see where the operator is, we use its commutator with the field at a spacetime point $y$, $\Phi_\tau(y)=i[\phi(y),O(\tau)]$. It is a classical solution of the Klein--Gordon equation, whose data at $x^0=0$ are $-\gamma_{2,\tau}$ and $\gamma_{1,\tau}$. In the causal diamond $D(A)$ it contains all the information of $O(\tau)$, since $O(\tau)=\int_\Sigma(\phi\,\partial_n\Phi_\tau-\Phi_\tau\partial_n\phi)$ on any Cauchy surface $\Sigma$ of $D(A)$, and it vanishes where $\phi(y)$ commutes with $O(\tau)$: a point-like operator appears as two null beams that cross at its position (Fig.~\ref{fig:flow}). We obtain $\Phi_\tau$ in $D(A)$ from Riemann's formula \cite{Courant:1962} and plot $\varepsilon=T(\xi,\xi)/4\pi^2$,
\begin{equation}
 \varepsilon=(\partial_{\omega_u}\Phi_\tau)^2+(\partial_{\omega_v}\Phi_\tau)^2+\tfrac12u(1-u)v(1-v)\,t^2\Phi_\tau^2,
 \label{eq:eps}
\end{equation}
where $T_{\mu\nu}$ is the stress tensor of $\Phi_\tau$, $u,v=x\mp x^0$, $\omega_u=\omega(u)$, $\omega_v=\omega(v)$, and 
\begin{equation}
\xi=2\pi[(x(1-x)-(x^0)^2)\partial_0+x^0(1-2x)\partial_x]\nonumber
\end{equation}
is the conformal Killing vector of $D(A)$. Without mass the flow is the geometric map $\omega_u\to\omega_u-2\pi\tau$, $\omega_v\to\omega_v+2\pi\tau$ generated by $\xi$ \cite{Hislop:1981uh}, and $\varepsilon$ is carried rigidly along its orbits, which go from the past to the future tip of $D(A)$.

\begin{figure}[t]
\centering\includegraphics[width=\linewidth]{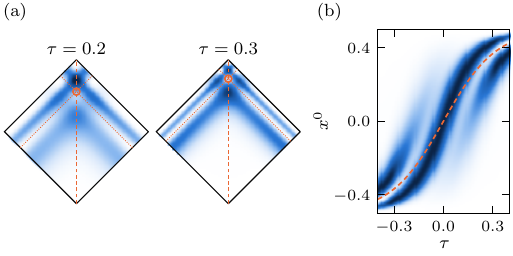}
\caption{The operator of Fig.~\ref{fig:flow} for $mL=3$. (a) $\varepsilon$ at $\tau=0.2$ and $0.3$: each beam splits in two. (b) World line on the axis: a second crossing, behind the conformal one (dashed), appears for $\tau\gtrsim0.2$.}
\label{fig:split}
\end{figure}

Fig.~\ref{fig:flow} shows the flow of an operator localized at the center of the interval. Without mass the two beams cross at the conformal position, $x^0=\frac12\tanh(\pi\tau)$, and keep their brightness. For $mL=1$ the operator runs ahead: at $\tau=0.2$ the beams cross at $x^0\simeq0.32$ instead of $0.28$, and they leave the null lines of the conformal flow. The world line on the axis is steeper than the conformal one [Fig.~\ref{fig:flow}(b)]. The effect grows with the mass. For $mL=3$ each beam splits in two, and the operator develops two crossings, one ahead of and one behind the conformal trajectory (Fig.~\ref{fig:split}): it is no longer localized near a point. At fixed mass, wider operators, which have more weight at small $s$, move further ahead, while narrow ones stay closer to the conformal trajectory (we checked this for $mL=3$). This has a simple reading in terms of the eigenvectors. By \eqref{eq:nsUV}, at large $s$ the modes are plane waves in $\omega$ with group velocity $2\pi$, as without mass. At small $s$ they are not: already the Rindler function $K_{is}(tx)$ oscillates as $x^{\pm is}$ only for $tx\lesssim s$, and the reflections encoded in $q_s$ also matter. The massive flow is thus dispersive: different modular frequencies travel differently, and this is how the nonlocal part of $\mH$ acts on localized operators. The smooth glow inside the light cone of the crossing is the massive wake, which is present also for local operators. For an operator off the center we also find that the beam moving towards the nearest end point stays sharp, as expected from the Bisognano--Wichmann limit, while the one crossing the middle of the interval spreads.

\textit{Discussion.}---We have obtained the modular Hamiltonian of a massive free field in a finite region, for any value of the mass. The key is the representation \eqref{eq:modes}--\eqref{eq:integral}, which solves the connection problem of the linear system associated with the Painlev\'e V transcendent \cite{Arias:2017dda}. It gives the eigenvectors as massive waves emitted from the end points and reflected between them, and makes the nonlocal structure of $\mH$ explicit. The eigenvectors also give the modular flow, which we have described in spacetime: the mass makes it dispersive, and a localized operator runs ahead of the conformal trajectory and spreads inside the causal diamond. 

The method uses only the existence of massive plane waves and of a Riemann--Hilbert problem in the rapidity plane, and it should extend to several intervals and to the massive Dirac field. Since the same function $S(z)$ is defined on the complement of the interval, where Green's identity gives $S\propto X\psi_s$ and $\partial_2S\propto P\chi_s$, it should also give the purification partners of the modular modes and the modular conjugation of the massive field. The explicit kernels allow exact computations of relative entropies and of other quantities that require more than the spectrum.

\begin{acknowledgments}
\textit{Acknowledgments}---We thank Horacio Casini, Marina Huerta and Pedro Martinez for reading a draft of this Letter. The authors are supported by CONICET and UNLP, Argentina. 
\end{acknowledgments}


\section*{End Matter}
\appendix

\section{Construction of the eigenvectors}
Let $Y(\kappa)$ be the $2\times2$ matrix, analytic in $\mathbb C-\mathbb R$, bounded at $\kappa=0$, with $Y\to\Id$ at infinity, and with jumps on the two half lines (oriented away from the origin, $\kappa>0$)
\begin{equation*}
 \begin{aligned}
 Y_+(\kappa)&=Y_-(\kappa)\big(\Id+2\sinh(\pi s)\,\kappa^{2is}d_t(\kappa)\,E_{12}\big),\\
 Y_+(-\kappa)&=Y_-(-\kappa)\big(\Id-2\sinh(\pi s)\,\kappa^{-2is}d_t(\kappa)\,E_{21}\big),
 \end{aligned}
\end{equation*}
where $E_{ij}$ are the matrix units, $(E_{ij})_{kl}=\delta_{ik}\delta_{jl}$, and $Y_\pm$ are the boundary values from the left ($+$) and from the right ($-$) of the oriented contour, i.e. from above on $\mathbb R_+$ and from below on $\mathbb R_-$. The jumps on $\mathbb R_+$ only change the second column of $Y$ and those on $\mathbb R_-$ the first one, and with $p_s(\kappa)=Y(\kappa)e_1$, $h_s(\kappa)=Y(-\kappa)e_2$ the Cauchy representation of the columns is \eqref{eq:integral}. For each row $(m_1,m_2)$ of $Y$ the density 
\begin{equation}
f(\kappa)=e^{\pm\pi s}m_1\kappa^{is}-m_2\,e^{\frac t2(\kappa+\kappa^{-1})}\kappa^{-is}\nonumber
\end{equation}
($\pm$ in the upper/lower half plane) has no jump, so that $\int_\Gamma\frac{\dd\kappa}\kappa e_\kappa(z)f(\kappa)$ solves the Klein--Gordon equation. Since $e_\kappa(z)e^{\frac t2(\kappa+\kappa^{-1})}=e_\kappa(z-1)$, its two terms are waves centered at $z=0$ and $z=1$. Outside $A$ both families can be rotated to a common half line and there is no cut. On $A$ they must be rotated to opposite half lines, and the factor $e^{\pm\pi s}$ produces exactly the jump $\lambda=e^{-2\pi s}$ of \eqref{eq:jump}; this gives \eqref{eq:modes}. The coefficient $\sinh(\pi s)/\pi$ in \eqref{eq:integral} cancels a constant term at each end point that is not allowed, so that $S$ is bounded and has only the two waves of \eqref{eq:tips}. The Carleman operator $1/(\kappa+\kappa')$ is diagonal in the Mellin transform, with eigenvalue $\pi/\cosh(\pi p)$, and $\kappa^{2is}$ shifts $p$ by $2s$; this bounds the geometric convergence rate by $\tanh^2(\pi s)$ per round trip. Since the jump matrices are unimodular, $\det Y$ is entire and tends to one, hence $\det Y=1$; the symmetries of the problem then fix the form of $\T_s=Y(0)$. After a diagonal gauge transformation the jumps do not depend on $\kappa$ and $t$, and the logarithmic derivatives of the solution with respect to $\kappa$ and $t$ form a Lax pair \cite{Jimbo:1981zz}, whose compatibility gives \eqref{eq:PV}. Replacing $q_s=1/d$ in \eqref{eq:PV} gives eq.~(6.5) of \cite{Arias:2017dda}, and its boundary conditions are our limits of small and large $t$.

The normalization follows from the end point behavior \eqref{eq:tips}
and $\int_0^1\dd x\,x^{iu-1}=\pi\delta(u)-i\,\mathrm{PV}(1/u)$, which give
\begin{equation*}
 \begin{aligned}
 &\int_A(\partial_2\Sv_{s'})^{+*}_i\Sv^+_{s,j}=(\G_s)_{ij}\,\delta(s-s'),\\
 &\G_s=\frac{\pi^2e^{-2\pi s}}{\sinh\pi s}\big[\Id+(\T_s\T_s^\dagger)^T\big],
 \end{aligned}
\end{equation*}
and the weights in \eqref{eq:kernels}.
Conservation of flux and Green's identity with the reciprocal Green
function, constructed from the sourced auxiliary problem
\eqref{eq:integral}, give uniqueness for prescribed incoming amplitudes.

Finally, we outline the proof of completeness of the eigenvectors. For $V(f,g)=(iPg,-iXf)$, self-adjoint in the vacuum covariance
inner product, the resolvent equation at $\xi$ reduces to
\eqref{eq:jump} with additive jumps and
$\lambda=(2\xi-1)/(2\xi+1)$. In the coordinate $\log[z/(1-z)]$,
each end has one open channel. Separating these waves from the
exponentially weighted remainder gives a Fredholm problem.
The flux and adjoint boundary pairings, together with end point
uniqueness, exclude its kernel and cokernel and give locally uniform
resolvent limits in the spectral interior. Their difference solves
the homogeneous cut problem. Stone's formula then gives the
absolutely continuous spectral measure in the two channels.
Nondegeneracy of the commutator excludes an atom of $(2V)^{-1}$ at
$0$, and antilocality of $\sqrt{-\partial_x^2+t^2}$ excludes atoms
at $\pm1$. This gives the full resolution of the identity. A detailed proof will be given in a longer paper.

\section{Checks}
We solve \eqref{eq:integral} with the trapezoidal rule in $\log\kappa$; for $s\gtrsim1$ extended precision is needed, because of cancellations of size $e^{\pi s}$ in \eqref{eq:modes}. The values of $q_s$ satisfy \eqref{eq:PV} and agree with a direct integration of the Painlev\'e equation to about ten digits. The sums $\sum_k\int\dd s\,\nu(s)\chi^k_s\chi^{k*}_s$ and $\sum_k\int\dd s\,\nu(s)\psi^k_s\psi^{k*}_s$, which by \eqref{eq:pair} and completeness must give $X$ and $P$, reproduce them with a relative error of $10^{-8}$. Wave packets of eigenvectors are orthonormal to $10^{-10}$, and Parseval's identity holds to $10^{-9}$--$10^{-12}$. On a harmonic chain the kernels can be computed without eigenvectors, $N=\int_0^1\dd\xi\,(P-\frac{\xi^2}4X^{-1})^{-1}$ and $M=\int_0^1\dd\xi\,(X-\frac{\xi^2}4P^{-1})^{-1}$. Applied to a smooth test function, \eqref{eq:kernels} agrees with a chain of $120$ sites to $4\times10^{-5}$ ($N_t$) and $10^{-3}$ ($M_t$), and to $3\times10^{-7}$ and $7\times10^{-5}$ after extrapolating in the number of sites. The modular flow of Fig.~\ref{fig:flow} agrees with the flow on the chain, $\exp[\tau(\begin{smallmatrix}0&-M\\ N&0\end{smallmatrix})]$, to $10^{-3}$ for $\tau\le0.2$, both for $mL=1$ and $mL=3$. The entropies (equation  \eqref{eq:Sn} below) agree with the chain to $10^{-9}$--$3\times10^{-8}$ (von Neumann). Eqs.~\eqref{eq:nsUV} and \eqref{eq:mirrorterm} come from a formal WKB expansion of the eigenvectors at fixed $t$, without uniform bounds on the remainder. Carried to order $s^{-3/2}$, it agrees with the spectral density of the exact eigenvectors up to $s=28$ for $mL=1$ and $s=20$ for $mL=2$, and reproduces the coefficient of the cusp to $2\%$. In $X$ the two terms of \eqref{eq:mirrorterm} give opposite singularities that cancel, as they must; conversely, this cancellation fixes the second term once the first is known.

\begin{figure}[t]
\centering\includegraphics[width=0.94\linewidth]{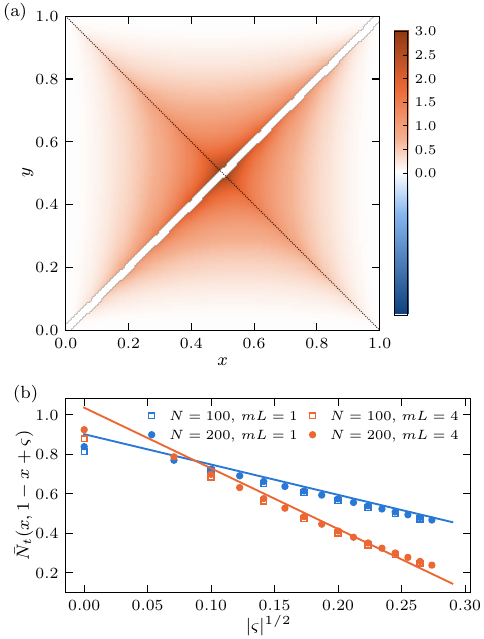}
\caption{(a) $N_t(x,y)$ for $mL=1$ on a chain of $200$ sites, computed without eigenvectors; a band $|x-y|\le3/200$ that contains the local term has been removed. Dotted: the mirror line $y=1-x$. (b) The average $\bar N_t$ of $N_t(x,1-x\pm\varsigma)$ at $x\simeq\frac14$, for $mL=1$ and $4$; the lines have the slope $-2\sqrt{\pi t x(1-x)}$ of \eqref{eq:mirror}.}
\label{fig:kernel}
\end{figure}

\section{Small mass}
With the equilibrium density $\varrho(x)=1/(\pi\sqrt{x(1-x)})$, whose logarithmic potential is constant on $A$, the Sherman--Morrison formula gives $X^{-1}=\mathcal Y+\frac{4\pi}\Lambda\varrho\varrho^T+O(t^2\log t)$, where $\mathcal Y$ is $t$ independent and annihilates constants. Inserting this in the formulas for $N$ and $M$ of the previous section gives \eqref{eq:small}, which the lattice confirms. Ref.~\cite{Baranov:2024rdm} finds the same form with coefficients four times smaller. That would violate the operator inequality $M_t\geq X^{-1}$, implied by the integral representation of the previous section, which \eqref{eq:small} saturates for test functions approaching a constant.

\section{Entropies and capacity}
The density of modes of the interval with a cutoff $\epsilon$ follows from \eqref{eq:tips},
\begin{equation*}
 \rho(s)=\tfrac2\pi\big[\log\tfrac2{m\epsilon}+\Rea\psi(is)\big]-\tfrac1\pi\partial_s\delta_s(t),
\end{equation*}
up to terms that oscillate in $s$. The first term is twice the density of the half line, and all the dependence on $L$ is in $\delta_s$. The entanglement and R\'enyi entropies only need this phase,
\begin{equation}
 \begin{aligned}
 S_n={}&\frac{n+1}{6n}\log\frac2{m\epsilon}+\kappa_n\\
 &+\frac n{n-1}\int_0^\infty\!\!\dd s\,\big[\coth(\pi ns)-\coth(\pi s)\big]\,\delta_s(t),
 \end{aligned}
 \label{eq:Sn}
\end{equation}
with $\kappa_n$ a nonuniversal constant. With \eqref{eq:PV} this reproduces the entropic $c$-functions of \cite{Casini:2005zv} to ten digits, the long distance term $-\frac18K_0(2mL)$ \cite{Cardy:2007mb}, and the small mass term $\frac12\log(-\log mL)$ \cite{Casini:2005zv}. In terms of the transmission probability $T_s=q_s^2/(1+q_s^2)$ between the end points, the $c$-functions are
\begin{equation*}
c_n=L\partial_LS_n=\frac{2n}{n-1}\int\dd s\,s[\coth\pi s-\coth\pi ns]\,T_s.
\end{equation*} The capacity of entanglement \cite{Yao:2010woi,deBoer:2018mzv}, 
\begin{equation}
C=\langle\mH^2\rangle-\langle\mH\rangle^2=\int\dd s\,\rho(s)\,\pi^2s^2/\sinh^2(\pi s),\nonumber 
\end{equation}
whose candidate $c$-function $L\partial_LC$ was studied in \cite{Arias:2023cap}, satisfies
\begin{equation*}
 L\partial_LC-c_1=2\pi\int_0^\infty\!\!\dd s\,\frac{s^3}{\sinh^2(\pi s)}\,\partial_sT_s .
\end{equation*}
The two $c$-functions differ only through the dependence of the transmission on the modular frequency. For $t\to0$ the right hand side is $1/(2\log\frac2t)$ and cancels the leading correction to $c_1$, so that $L\partial_LC=\frac13+O(\log^{-2}t)$, as found in \cite{Arias:2023cap}. At large $t$, $L\partial_LC\simeq\frac{\pi t}4e^{-2t}$, while $c_1\simeq\frac{\sqrt\pi}8\sqrt t\,e^{-2t}$.

\section{Modular flow in the diamond}
For $\gamma_2=0$ at $\tau=0$ the eigenvectors give
\begin{equation*}
 \begin{aligned}
 \gamma_{1,\tau}&=\sum_k\int_0^\infty\!\!\dd s\,\cos(2\pi s\tau)\,\psi^k_s\,\langle\chi^k_s,\gamma_{1,0}\rangle,\\
 \gamma_{2,\tau}&=\sum_k\int_0^\infty\!\!\dd s\,\sin(2\pi s\tau)\,\chi^k_s\,\langle\chi^k_s,\gamma_{1,0}\rangle.
 \end{aligned}
\end{equation*}
For $T=|x^0|$ and $r=\sqrt{T^2-(x-y)^2}$, Riemann's formula with data $f=\Phi_\tau|_{x^0=0}$, $g=\partial_0\Phi_\tau|_{x^0=0}$ reads
\begin{eqnarray}
\Phi_\tau&=&\frac12[f(x-T)+f(x+T)]\nonumber\\&&+\frac12\mathrm{sgn}(x^0)\int_{x-T}^{x+T}\dd y\,g(y)J_0(tr)\nonumber\\&&-\frac12tT\int_{x-T}^{x+T}\dd y\,f(y)\frac{J_1(tr)}r.\nonumber
\end{eqnarray}
The first term carries the beams, and the Bessel terms are the wake. In the modular coordinates $\lambda=\frac12(\omega_u+\omega_v)$, $\rho=\frac12(\omega_v-\omega_u)$ the diamond is the whole plane, 
\begin{equation}
\dd s^2=u(1-u)v(1-v)(\dd\lambda^2-\dd\rho^2),\nonumber
\end{equation}
and $\varepsilon$ in \eqref{eq:eps} is the energy density for the time $\rho$; the conformal flow is $\rho\to\rho+2\pi\tau$.

\end{document}